
\magnification=1200
\hsize=15truecm
\vsize=23truecm
\baselineskip 20 truept
\voffset=-0.5truecm
\parindent=0cm
\overfullrule=0pt

\def\Ai{\hbox{\hbox{${\cal A}$}}\kern-1.9mm{\hbox{${/}$}}}
\def\Vi{\hbox{\hbox{${\cal V}$}}\kern-1.9mm{\hbox{${/}$}}}
\def\Di{\hbox{\hbox{${\cal D}$}}\kern-1.9mm{\hbox{${/}$}}}
\def\lam{\hbox{\hbox{${\lambda}$}}\kern-1.6mm{\hbox{${/}$}}}
\def\D{\hbox{\hbox{${D}$}}\kern-1.9mm{\hbox{${/}$}}}
\def\A{\hbox{\hbox{${A}$}}\kern-1.8mm{\hbox{${/}$}}}
\def\V{\hbox{\hbox{${V}$}}\kern-1.9mm{\hbox{${/}$}}}
\def\parz{\hbox{\hbox{${\partial}$}}\kern-1.7mm{\hbox{${/}$}}}
\def\B{\hbox{\hbox{${B}$}}\kern-1.7mm{\hbox{${/}$}}}
\def\R{\hbox{\hbox{${R}$}}\kern-1.7mm{\hbox{${/}$}}}
\def\si{\hbox{\hbox{${\xi}$}}\kern-1.7mm{\hbox{${/}$}}}
\def\slash#1{\setbox0=\hbox{$#1$}#1\hskip-\wd0\dimen0=5pt\advance
\dimen0 by - \ht0\advance\dimen0 by\dp0\lower0.5\dimen0\hbox
to\wd0{\hss\sl/\/\hss}}

\qquad
The principal advantage of the Green--Schwarz heterotic string sigma model
with respect to the Neveu--Schwarz--Ramond formulation is its
manifest N=1, D=10 space--time supersymmetry. It is this property which
renders it particularly suitable for the derivation of the low energy
effective superstring theory, i.e. N=1, D=10
Supergravity--Super--Yang--Mills, in superspace. Indeed, the
$\kappa$--anomaly  cancellation mechanism [1] constitutes a
systematic approach for the derivation of superspace constraints
for the low energy effective theory which are automatically consistent with
the Bianchi identities in superspace: the Wess--Zumino consistency
condition on the $\kappa$--anomalies ensures that they can be cancelled
by imposing suitable constraints on the superfields of the low
energy theory and that with these constraints the Bianchi identities
can be consistently solved.

\qquad
Along these lines in ref. [2], see also [3], the order $\alpha'$ one--loop
$\kappa$--anomalies of the heterotic string sigma--model have been
explicitly determined, by a direct perturbative computation, together
with the order $\alpha'$ superspace constraints which give rise to
their cancellation.

\qquad
On the other hand, exact solutions, i.e. to all orders in $\alpha'$, of the
relevant Bianchi identities have been found previously in the literature
[4,5,6].

\qquad
In this letter we point out that the exact constraints found in this way,
if truncated to order $\alpha'$, appear to differ from the
ones found through the $\kappa$--anomaly cancellation mechanism in ref.
[2], the difference not being simply related to the usual ambiguity in the
choice of standard constraints for a theory formulated in superspace. We
present the solution of this puzzle by showing that the difference between
the two sets of constraints is related on one hand to a
{\it trivial}  $\kappa$--anomaly and that, on the other hand,
this difference can be eliminated by order--$\alpha'$ superfield
redefinitions.

\qquad
The heterotic string sigma model action is given by

$$
I = - {1\over 2} \int d^2 \sigma \sqrt{g} \left(V^a_- V_{+a}+ V_-^A V_+^B
B_{BA} - \psi {\cal D}_- \psi \right) \eqno(1)
$$

where the induced Zehnbein are given by $V_i^A = \partial_i Z^M E_M{}^A(Z)$
and the index $A=(a, \alpha)$ stands for ten bosonic (a=1, ..., 10) and
sixteen fermionic $(\alpha=1, ..., 16)$ entries. The two--dimensional
light--cone indices are introduced via $V_\pm^A = e_\pm^i V_i^A$ (see [2]
for the notation). The components of the two--super form $B$ are given in
Zehnbein basis by $B = {1\over 2} E^A E^C B_{CA}$ where $E^A = dZ^M
E_M{}^A (Z)$.

\qquad
A p--superform $\wedge^p$ can be decomposed in sectors with $r$ bosonic
$(E^a)$ and $s$ fermionic $(E^\alpha)$ Zehnbein according to

$$
\wedge^p = {1 \over p!}  E^{A_1} \cdots E^{A_p} \wedge_{A_p}..._{A_1} =
\sum_{r+s=p} \wedge_{(r,s)}. \eqno(2)
$$

For what follows it is convenient to introduce the
following subspace ${\cal H}_4$ of closed four--superforms:

$$
{\cal H}_4 = \{\wedge^4 |d \wedge^4 =0, \wedge_{(0,4)} = 0 = \wedge_{(1,3)} \}.
\eqno(3)
$$

The heterotic fermions $\psi$ stay in the fundamental representation of
$SO(32)$ and ${\cal D}_- = e^i_- (\partial_i - V_i^B A_B)$ where $A_B$ are the
components of the connection one--superform $A=E^B A_B$, with values
in the Lie algebra of $SO(32)$.

\qquad
The action (1) is invariant under $\kappa$--transformations, with
transformation parameter $\kappa_{+ \beta}$, a space--time spinor, which are
induced by

$$\eqalign{
& \delta_k Z^M = \Delta^\alpha E_\alpha^M \cr
& \delta_\kappa \psi = \Delta^\alpha A_\alpha \psi \cr
& \Delta^\alpha \equiv V_-^a (\Gamma_a)^{\alpha \beta} \kappa_{+ \beta} =
(\slash V_- \kappa_+)^\alpha\cr}. \eqno(4)
$$

Due to the Virasoro constraint, $V^2_- = V_-^a V_{-a} =0,$ we have
that $\slash V_- \Delta = V_-^2 \kappa_+ =0$. To be more precise, taking the
Virasoro constraint into account\footnote*{The use of the Virasoro constraint
can be avoided by introducing non trivial $\kappa$--transformations
for the zweibein $e^i_+$. Here, to avoid a merely technical
complication, we prefer to enforce the equation of motion for the
world--sheet metric, i.e. $V^2_-=0$.}, under the transformations (4) the
action varies as follows

$$
\delta_\kappa I = -{1\over 2} \int d^2 \sigma \sqrt{g} \left(2 V_-^\beta
\slash V_{+\beta\gamma} \Delta^\gamma + V_-^C V_+^D \Delta^\gamma (d B)_{\gamma
DC} \right). \eqno(5)
$$

To get (5) we used the standard superspace constraints on the torsion,
$T^A = DE^A = dE^A + E^B \Omega_B{}^A = {1\over 2} E^B E^C T_{CB}{}^A$ where
$\Omega_A{}^B$ is the $D=10$ Lorentz connection one--form,

$$\eqalign{
& T_{\alpha\beta}{}^a = 2 {\Gamma^a}_{\alpha\beta} \cr
& T_{\alpha a}{}^b = 0, \cr} \eqno(6)
$$

and on the Yang--Mills curvature two form $F= d A+AA = {1 \over 2} E^A E^B
F_{BA},$

$$
F_{\alpha \beta} =0. \eqno(7)
$$

(5) can now be set to zero by choosing standard  constraints for $H^0
\equiv  d B$,

$$\eqalign{
& H^0_{\alpha \beta \gamma} = H^0_{a b \alpha} =0 \cr
& H^0_{a \alpha \beta} = 2 (\Gamma_a)_{\alpha \beta}. \cr}
\eqno(8)
$$

\qquad
At the quantum level the presence of $\kappa$--anomalies may require to
modify the constraints (6),(7) and (8). Indeed, on general grounds [1]
the $\kappa$--anomaly $A_\kappa$ is structured in three terms,
${\cal A}_\kappa = {\cal A} + {\cal A}_T + {\cal A}_F; {\cal A}_T$
and ${\cal A}_F$ induce quantum corrections to (6) and (7) respectively
which are expected to be absent, at least at one loop i.e.
at order $\alpha'$, while
${\cal A}$ induces corrections to (8) which appear already at order
$\alpha'$. Its general one--loop structure is given by\footnote{**}{Our
normalizations are determined by defining the effective action as
$exp ({i \over 2\pi \alpha'} \Gamma) = \int \{{\cal D} \varphi\} exp
({i \over 2\pi \alpha'} I)$, $\Gamma = I + \alpha' \Gamma^{(1)} +
o (\alpha'^2), {\cal A}_\kappa = \alpha' \delta_\kappa \Gamma^{(1)}$.}

$$
{\cal A}= - {\alpha' \over 8} \int d^2 \sigma \sqrt{g} V_-^A V_+^B
\Delta^\gamma \left(\omega_{3 Y M} - \omega_{3L} + X \right)_{\gamma BA}
\eqno(9)
$$

where $\omega_{3 YM}$ and $\omega_{3L}$ are respectively the Yang--Mills and
Lorentz Chern--Simons three superforms satisfying $d \omega_{3YM} = tr FF, d
\omega_{3L} = tr RR$, while $X_{CBA}$ are the components of a gauge
and Lorentz--invariant three superform $X = {1\over 3!} E^A E^B E^C
X_{CBA}$.

The Wess--Zumino consistency condition on (9) becomes

$$
\delta_\kappa {\cal A} = - {\alpha' \over 8} \int d^2 \sigma \sqrt{g} V^C_-
V_+^D \Delta^\alpha \Delta^\beta \Bigl(tr FF - tr RR + dX\Bigr)_{\beta\alpha
DC} =0,  \eqno(10)
$$

which is precisely equivalent to the condition

$$
\alpha' (tr FF - tr RR + dX) \in {\cal H}_4.
$$

Since we have an $\alpha'$ in front, in checking this we can use the
classical constraints (6), (7); moreover, following [7] we can impose the
classical constraint $R_{\alpha \beta a}{}^b =0$ on the Lorentz curvature
two--super form $R_a{}^b = {1\over 2} E^C E^D R_{DCa}{}^b=$ $d \Omega_a{}^b
+ \Omega_a{}^c \Omega_c{}^b$. As a conseguence of
$F_{\alpha\beta}=0$ and $R_{\alpha \beta a}{}^b =0$ one has that $tr FF
\in {\cal H}_4$ and $tr RR \in {\cal H}_4$ separately.
One remains with the condition

$$
d X \in {\cal H}_4 \eqno(11)
$$

at zero--order in $\alpha'$.

The anomaly can now be cancelled by imposing

$$
\delta_\kappa \Gamma = \delta_\kappa I + {\cal A} = 0 \eqno(12)
$$

which amounts to define the generalized $B-$ curvature

$$
H= d B + {\alpha' \over 4} (\omega_{3YM} - \omega_{3L}) \eqno(13)
$$

and to impose on it the constraints

$$\eqalign{
& H_{\alpha \beta\gamma} = -{\alpha'\over 4} X_{\alpha \beta \gamma} \cr
& H_{a \alpha \beta} = 2(\Gamma_a)_{\alpha\beta} - {\alpha'\over 4} X_{a
\alpha \beta} \cr
& H_{a b \alpha} = -{\alpha' \over 4} X_{a b \alpha}. \cr} \eqno(14)
$$

Thanks to (11) the Bianchi identity associated to
(13), i.e.

$$
dH = {\alpha'\over 4} \left(tr FF - tr RR \right), \eqno(15)
$$

can be consistently solved with the constraints (14).

\qquad
An {\it explicit} computation of the one--loop $\kappa$--anomaly
has been performed in [2] and the result was indeed formula (9), but
with $X=0$, meaning that the order $\alpha'$ corrections to the H--field
in (14) are absent.

\qquad
On the other hand, exact solutions of the Bianchi identity (15), based on
the  Bonora--Pasti--Tonin (BPT) theorem [4], are known in the literature [5],
[6]. These solutions differ among themselves only because of different
choices for classical standard constraints, such as (6) and (8).
The choice of ref. [6] differs from ours just by a shift of the Lorentz
connection $(\Omega_a{}^b \rightarrow \Omega_a{}^b + E^c T_{ca}{}^b$, where
$T_{cab}$ is the vectorial torsion) and so we make our comparison with that
reference.

\qquad
The BPT theorem ensures that one can write $tr RR = d \tilde X + K$ where
$\tilde X$ is an invariant three superform and $K \in {\cal H}_4$. Then
(15) can be written as

$$
d \left(H+ {\alpha' \over 4} \tilde X \right)= {\alpha' \over 4}
\left(tr F^2 - K \right) \in {\cal H}_4 \eqno(16)
$$

and the {\it exact} solution is found by imposing on H the constraints
in (14) with $X \rightarrow \tilde X$. However, the $\tilde X$ found in
[6], when truncated to zero--order in $\alpha'$, is different from zero.
Calling $X = \tilde X |_{\alpha'=0}$ the authors of ref. [6] got

$$\eqalign{
& X_{\alpha \beta \gamma}=0 \cr
& X_{a\alpha \beta} = -{1 \over 3} (\Gamma_a{}^{c_1-c_4})_{\alpha \beta}
(R_{c_1-c_4} + T_{c_1 c_2}{}^\gamma (\Gamma_{c_3 c_4})_\gamma{}^\delta \lambda_
\delta) \cr} \eqno(17)
$$

where $T_{ab}{}^\alpha$ is the gravitino field strength and $\lambda_\beta$
is the gravitello. We  do not report the explicit expressions of $X_{(2,1)}$
and  $X_{(3,0)}$ since they are completely  determined by (17) and
by the condition

$$
dX \in {\cal H}_4  \ ({\rm at \ zero-order \ in} \  \alpha'). \eqno(18)
$$

Eq. (17) seems thus to be in contrast with the result, $X=0,$ of ref. [2].

\qquad
To solve this puzzle we translate first the expression (17) in the
``missing" contribution to the anomaly

$$
{\cal A}_X = -{\alpha'\over 8} \int d^2 \sigma \sqrt{g} V_-^A V_+^B
\Delta^\gamma X_{\gamma BA} \eqno(19)
$$

and note then that the three form $X$, determined through (17) and (18),
admits the following remarkable decomposition at zero--order in $\alpha'$
(meaning that one can use the equations of pure supergravity to verify it):

$$
X= d C + X^{(0)} \eqno(20)
$$

where the two--form $C$ is given by

$$\eqalign{
& C_{\alpha \beta} =0 \cr
& C_{a \alpha} = -{2\over 3} (\Gamma_a \Gamma^{bc})_{\alpha \beta}
T_{bc}{}^\beta \cr
& C_{ab} = -{4 \over 3} R_{[ab]} - {1\over 3} T_{cd}{}^\alpha
(\Gamma_{ab}{}^{cd})_\alpha{}^\beta \lambda_\beta \cr} \eqno(21)
$$

and the three form $X^{(0)}$ is given by

$$\eqalign{
& X^{(0)}_{\alpha \beta \gamma} =0 \cr
& X^{(0)}_{a\alpha \beta} = - {2\over 3} R (\Gamma_a)_{\alpha\beta} \cr
& X^{(0)}_{ab \alpha} = {1 \over 3} (\Gamma_{ab})_\alpha{}^\beta D_\beta R \cr
& X^{(0)}_{abc} = -{1 \over 3} R T_{abc} + {1\over 8}
(\Gamma_{abc})^{\alpha\beta} \left(\lambda_\alpha D_\beta R - {1\over 3}
D_\alpha D_\beta R \right).\cr} \eqno(22)
$$

Here $R_{ab} \equiv R_a{}^c{}_{bc}$ and $R\equiv R_a{}^a$. Moreover

$$
d X^{(0)} \in {\cal H}_4. \eqno(23)
$$

Notice that the property (23) holds actually {\it exactly} i.e. to all
orders in $\alpha'$, due to the definition (22).

Inserting (20) and (22) in (19) we get for the ``missing" anomaly

$$
{\cal A}_X = -{\alpha' \over 8} \int d^2 \sigma \sqrt{g} \left(V_-^C V_+^D
\Delta^\gamma (dC)_{\gamma DC} + {1\over 3} V_-^a V_{+a} \Delta^\alpha
D_\alpha R + {2 \over 3} R V^\alpha_- \slash V_{+\alpha\gamma} \Delta^\gamma
\right),
$$

and it is not difficult to realize that this is actually a trivial
cocycle

$$
{\cal A}_X  = \delta_\kappa \left(-{\alpha' \over 8} \int d^2 \sigma \sqrt{g}
\left(V_-^A V^B_+ C_{BA} + {1\over 3} V_-^a V_{+a} R \right) \right) \eqno(24)
$$

which can be eliminated by subtracting from the classical action a local
counter term

$$
I \rightarrow I + {\alpha' \over 8} \int d^2 \sigma \sqrt{g} \left(V_-^A V_+^B
C_{BA} + {1\over 3} V_-^a V_{+a} R \right). \eqno(25)
$$

Equivalently this trivial ``anomaly" can be eliminated by the redefinitions

$$\eqalign{
& B^* = B + {\alpha' \over 4} C \cr
& E^{a*} = \left(1+ {\alpha'\over 24} R \right)E^a. \cr} \eqno(26)
$$

This explains that in ref. [2] no non--trivial one--loop
anomaly has been lost.

\qquad
On the other side the exact solution of ref. [6] is based on the identity
(BPT theorem)

$$
tr RR = d \tilde X + K \eqno(27)
$$

where $K \in {\cal H}_4$ exactly. However, this decomposition is not
unique for two reasons: first, $\tilde X$ is defined only modulo
exact forms and, second, if we find a three form $Z$ such
that $d Z \in {\cal H}_4$ the decomposition (27)
holds also if we replace $\tilde X \rightarrow \tilde X - Z$ and $K
\rightarrow K + d Z \in {\cal H}_4$.
Now, since (20)  holds at zero order we can write the exact relation

$$
\tilde X = d C + X^{(0)} + \alpha' X^{(1)} \eqno(28)
$$

for some three--form $X^{(1)}$. Substituting this in (27) we get

$$
tr RR = d \left(\alpha' X^{(1)} \right) + K', \eqno(29)
$$

where $K'=K+dX^{(0)} \in {\cal H}_4$ due to the fact that (23)
holds exactly. With (29) the Bianchi identity (15) becomes

$$
d \left(H+ {\alpha'^2 \over 4} X^{(1)} \right) = {\alpha' \over 4}
\left(tr FF - K' \right) \in
{\cal H}_4 \eqno(30)
$$

meaning that with this decomposition there are no order--$\alpha'$
corrections to the classical superspace constraints of $H$, see
(8), (they start at order $\alpha'^2$) in complete agreement with
the results from the $\kappa$--anomaly cancellation mechanism.

\qquad
Equivalently, the solution of ref. [6] can be transformed to
a solution where there are no order--$\alpha'$ corrections to the
classical $H$ constraints by accompanying the redefinitions (26)
with

$$\eqalign{
& E^{* \alpha}= \left(1+ {\alpha' \over 48} R \right)
E^\alpha - {\alpha'\over 48}
\Bigl((\Gamma_a)^{\alpha \gamma} D_\gamma R \Bigr) E^a \cr
& \Omega^{*b}_a = \Omega_a{}^b - {\alpha' \over 24} E^\alpha
(\Gamma_{ab})_\alpha{}^\beta D_\beta R.\cr} \eqno(31)
$$

These redefinitions, apart from eliminating $X^{(0)}$ and $dC$ from the
$H-$ constraints, are just the ones which preserve (6).

\qquad
The conclusions of this letter are that there is perfect agreement between
results from the $\kappa$--anomaly cancellation mechanism and
exact solutions to the $H-$Bianchi identities determined previously, and
that there exists an exact solution which at first order in $\alpha'$
becomes rather simple.
At first order in $\alpha'$ one has in particular

$$\eqalign{
& T_{\alpha \beta}{}^a = 2 \Gamma^a_{\alpha \beta} \cr
& T_{\alpha\beta}{}^\gamma = 2 \delta^\gamma_{(\alpha} \lambda_{\beta)} -
\Gamma^g_{\alpha\beta} (\Gamma_g)^{\gamma \epsilon} \lambda_\epsilon \cr
& T_{\alpha a}{}^b = 0 \cr
& T_{a \alpha}{}^\beta = {1 \over 4} (\Gamma^{bc})_\alpha{}^\beta T_{abc} -
{\alpha' \over 8} (\Gamma_a)_{\alpha \epsilon} \Bigl(tr (\chi^\epsilon
\chi^\beta) - tr (T^\epsilon T^\beta) \Bigr) \cr
& R_{\alpha\beta ab} = -{\alpha'\over 4} (\Gamma_{[a})_{\alpha\epsilon}
\Bigl(tr(\chi^\epsilon \chi^\varphi) - tr (T^\epsilon T^\varphi) \Bigr)
(\Gamma_{b]})_{\varphi \beta} \cr
& R_{a \alpha bc}= 2(\Gamma_a)_{\alpha\beta} T_{bc}{}^\beta + {3 \alpha'
\over 8} (\Gamma_{[a})_{\alpha\beta} \Bigl(tr(F_{bc]} \chi^\beta) - tr
(R_{bc]} T^\beta) \Bigr)\cr
& F_{\alpha\beta}=0 \cr
& F_{a \alpha} = 2(\Gamma_a)_{\alpha\beta} \chi^\beta \cr
& H_{\alpha \beta \gamma} =0 \cr
& H_{a \alpha\beta} =2 (\Gamma_a)_{\alpha\beta} \cr
& H_{a b \alpha} =0 \cr
& H_{a b c} = T_{abc} \cr} \eqno(32)
$$

Here $\chi^\alpha$ is the gluino superfield and
$tr(T^\alpha T^\beta) = T_{ab}{}^\alpha T^{b a \beta}$ etc.
The dilaton $\phi$ satisfies $\lambda_\alpha= D_\alpha \phi$.

\qquad
Finally we would like to notice that, since with the above
parametrizations the Bianchi identities close only up to order $\alpha'$,
the $X^{(1)}$ appearing in (29) is clearly non vanishing and there has to
be a {\it non trivial} contribution of the $X-$type, see (9), to the
{\it two--loop} $\kappa-$anomaly.

\qquad
Moreover, the so called ``poltergeister", appearing in N=1, D=10
supergravity theory when the Lorentz--Chern--Simons term is present,
which in previous formulations [4,5,6] show up at order $\alpha'$, in the
formulation corresponding to (32) are shifted to order $(\alpha')^2$.

\vfill\eject

{\bf References}
\vskip 0.5truecm
[1] M. Tonin, Int. J. Mod. Phys. A3 (1988) 1519; A4 (1989) 1983; A6 (1991)
315.
\smallskip
[2] A. Candiello, K. Lechner and M. Tonin, Nucl. Phys. B438, (1995) 67.
\smallskip
[3] I.I. Atick, A. Dhar and B. Ratra, Phys. Lett. B169 (1986) 54;

M.T. Grisaru, H. Nishino and D. Zanon, Nucl. Phys. B314 (1989) 363.
\smallskip
[4] L. Bonora, P. Pasti and M. Tonin, Phys. Lett. B188 (1987) 335;
\smallskip
[5] R. D'Auria, P. Fr\`e, M. Raciti and F. Riva, Int. J. Mod. Phys. A3 (1988)
953;

L. Castellani, R. D'Auria and P. Fr\`e, Phys. Lett. B196 (1987) 349;

L. Bonora, M. Bregola, K. Lechner, P. Pasti and M. Tonin, Nucl. Phys. B296
(1988) 877.
\smallskip
[6] M. Raciti, F. Riva and D. Zanon, Phys. Lett. B227 (1989) 118.
\smallskip
[7] L. Bonora, P. Pasti and M. Tonin, Nucl. Phys. B286 (1987) 150;

A. Candiello and K. Lechner, Nucl. Phys. B412 (1994) 479.

\vfill\eject
\nopagenumbers

\rightline{DFPD/95/TH/32}

\vskip 0.3truecm
\rightline{May 1995}
\vskip 1truecm

\centerline{\bf String $\kappa$--anomalies and $D=10$ Supergravity}
\centerline{\bf constraints: the solution of a puzzle}
\vskip 0.5truecm
\centerline{\bf K. LECHNER$^*$}

\vskip 0.3truecm
\centerline{\bf Dipartimento di Fisica, Universit\`a di Padova}
\centerline{\bf and}
\vskip 0.3truecm
\centerline{\bf Istituto Nazionale di Fisica Nucleare, Sezione di Padova}
\centerline{\bf Italy}

\vskip 2.5truecm
\centerline{\bf Abstract}
\vskip 0.3truecm
The $\kappa$--anomaly cancellation mechanism in the heterotic superstring
determines the superspace constraints  for N=1, D=10
Supergravity--Super--Yang--Mills theory. We point out that the constraints
found recently in
this way appear to disagree with superspace solutions found in the past.
We solve this puzzle establishing perfect agreement between the two
methods.

\vskip 4truecm

*Supported
in part by M.P.I. This work is carried out in the framework of the European
Community Programme ``Gauge Theories, Applied Supersymmetry and Quantum
Gravity" with a financial contribution under contract SC1--CT92--D789.

\bye